\documentclass[prb,twocolumn,amsmath,amssymb,showpacs]{revtex4}
\usepackage{graphicx}

\def\beq{\begin{equation}}
\def\eeq{\end{equation}}
\def\beqa{\begin{eqnarray}}
\def\eeqa{\end{eqnarray}}
\def\rmd{{\rm d}}

\begin{document}

\title{Friedel oscillations in a two-band Hubbard model for 
CuO chains}

\author{M. Mori$^{1}$, T. Tohyama$^{1}$. S. Maekawa$^{1}$, and
J. A. Riera$^{1,2,3}$} 
\affiliation{
$^{1}$Institute for Materials Research, Tohoku University, Sendai 980-8577, 
Japan \\
$^{2}$Instituto de F\'{\i}sica Rosario, Consejo Nacional de
Investigaciones
Cient\'{\i}ficas y T\'ecnicas, y Departamento de F\'{\i}sica,\\
Universidad Nacional de Rosario, Avenida Pellegrini 250,
2000-Rosario, Argentina \\
$^{3}$Laboratoire de Physique Th\'eorique CNRS-FRE2603,
Universit\'e Paul Sabatier, F-31062 Toulouse, France}

\date{\today}

\begin{abstract}
Friedel oscillations induced by open boundary conditions in 
a two-band Hubbard model for CuO chains are numerically studied.
We find that for physically realistic parameters and close to
quarter filling, these oscillations have a $2k_F$ modulation according
with experimental results on YBa$_2$Cu$_3$O$_{7-\delta}$. In 
addition, we predict that, for the same parameters, as hole doping
is reduced from quarter filling to half filling, Friedel oscillations
would acquire a $4k_F$ modulation, typical of a strongly correlated
electrons regime. The $4k_F$ modulation dominates also in the
electron doped region. 
The range of parameters varied is very broad, and hence the results
reported could apply to other cuprates and other strongly correlated
compounds with quasi-one dimensional structures. On a more theoretical
side, we stress the fact that the copper and oxygen subsystems should
be described by two different Luttinger liquid exponents.

\end{abstract}


\maketitle

\section{Introduction}
\label{Intro}

The application of scanning tunneling spectroscopy (STM) techniques
to the CuO chain planes in YBa$_2$Cu$_3$O$_{7-\delta}$ (YBCO) has provided
new insights on the physics of these quasi-one dimensional (1D) electronic 
structures.
These experimental studies have shown in real space the presence of
charge modulations along the chains at low temperature, inside the
superconducting phase.\cite{derro,maki}
These charge oscillations had been inferred earlier from neutron
scattering experiments on the same compound.\cite{mook}
Taking together results from both band calculations\cite{pickett} and
angle-resolved photoemission\cite{lu} (ARPES) indicating a chain
filling close to one quarter,
it turns out that the observed charge modulations have a $2k_F$
wavenumber.

Although earlier STM studies\cite{edwards} had given indications of
the presence of charge oscillations on the chains, the present
interest on these features comes from a more recent set of
experiments exploring the interplay between the chain plane and the
CuO$_2$ plane in YBCO. On one side, the chains (running along the
{\bf b} direction) may contribute to the in-plane resistivity
anisotropy in this compound.\cite{ando} It is also important
to notice that stripes in CuO$_2$ planes also run along the {\bf b}
direction, may be not coincidentally. On the other side, there have
been experimental indications of a superfluid density induced on the
chains due to the proximity to the CuO$_2$
planes.\cite{basov,gagnon} Although the interpretation of this 
last set of experiments is still controversial\cite{grevin,lu}, one
recent theoretical study\cite{morr} offers an explanation of STM
results\cite{derro} based on this scenario of proximity induced
chain superconductivity (SC). In this model, the resonances in the
chains arise from the interference between magnetic impurities in
the chains.

Alternatively, we believe it is necessary to search for explanations
of these experimental results based on models which capture the
intrinsic electronic interactions on the chains. Eventually, in
this kind of models, the SC on the CuO$_2$ planes could modify
at an effective level the coupling constants of the intrachain
interactions. This approach connects the present study to the more
general theoretical interest in various types of charge
inhomogeneities appearing in strongly correlated electron systems
in low spatial dimensions. In fact, in addition to the relevance of
the presence of charge oscillations in the chains to the physics of
the CuO$_2$ planes, it has been emphasized that precisely the assumed
strong Coulomb interactions in the cuprates would naively favour
a $4k_F$ instead of the observed $2k_F$ wavenumber of the charge 
modulations.\cite{maki}

Since CuO chains are cut by oxygen depletion, we will try to describe
the charge modulations as generalized Friedel oscillations (FO) starting
at the open end of the resulting fragments. A very recent numerical
and analytical study\cite{white_friedel} on the 1D one-band Hubbard 
model has
indeed found a crossover from $2k_F$ to $4k_F$ FO
as the strength of the onsite Coulomb interaction $U$ is increased.
We are going to examine a two-band Hubbard model, appropriate for 
the CuO  chains, where the situation is more complex due to
different Coulomb repulsions on Cu and O ions, in addition to the
charge transfer between them.\cite{pencmila} It is interesting to
notice that ARPES shows\cite{lu} a chain's Fermi surface in agreement
with band calculations and at the same time the dispersion along the
chain direction agrees with the holon band predicted by strongly
correlated theories.\cite{maekawa}
Hence, although the Coulomb repulsion on Cu ions is large, it does
not automatically imply a strongly correlated behavior on any given
physical quantity. In particular, we will show a complex dependence
of $2k_F$ and $4k_F$ FO with respect to the Coulomb interactions
and charge transfer parameter.

The model here studied does not include electron-lattice coupling
which is another possibility of inducing $2k_F$ instabilities.
Experimentally\cite{mook}, it has been suggested that standard
lattice phonons are unlikely to reproduce the observed features.
Finally, both in order to help the understanding of our results and
in order to extend the scope of this study, we will consider a
somewhat large variation of interaction parameters, and in
addition  to quarter-filling, dopings close to half-filling, and
also electron doped chains will also be studied. Besides, we will
consider open ends occupied with O ions and with Cu ions.

This paper is structured as follows. In Section \ref{phasediag}, we
describe the model and method of calculation, and we show the 
numerical results for density oscillations. The theoretical and
experimental implications of these results are discussed in
Section \ref{discussion}.

\section{$2k_F$/$4k_F$ phase boundary}
\label{phasediag}

\subsection{Model and method of calculation}

The model studied is the 1D two-band Hubbard model defined as:
\begin{eqnarray}
{\cal H} = &-& t_{pd} \sum_{i,\sigma }
(c^{\dagger}_{ 2i-1 \sigma}
c_{ 2i \sigma} + h.c. )  \nonumber \\
&+& U_d \sum_{i} n_{2i-1 \uparrow} n_{2i-1 \downarrow} +
U_p \sum_{i} n_{2i \uparrow} n_{2i \downarrow}  \nonumber \\
&+& U_{pd} \sum_{i} n_{2i-1} n_{2i} +
\Delta \sum_{i} n_{2i}
\label{hamhub}
\end{eqnarray}
\noindent
where, $c^{\dagger}_{j\sigma}$ creates a hole with spin $\sigma$ at
site $j$, $n_{j\sigma}=c^{\dagger}_{j\sigma} c_{j\sigma}$,
$n_{j}=n_{j \uparrow} + n_{j \downarrow}$. The sums run over
$i=1,\ldots,L$, $L$: number of unit cells, and Cu (O) ions
occupy odd (even) sites. $t_{pd}=1$. All our results shown below
correspond to $U_d=8$, $U_{pd}=0$, and $U_p, \Delta < U_d$. The
number of holes ($N_h$) is larger than $L$ for hole doped systems
in which case the filling is computed as $n_h= (2L-N_h)/L$, or
smaller than $L$ for electron doped systems in which case the
filling is computed as $n_e= N_h/L$. At half filling $n_h=n_e=1$. The 
numerical technique employed in the present study is the density
matrix renormalization group algorithm (DMRG).\cite{white-DMRG} The
most important quantity measured by this technique is the on-site
charge density, $n(i)=\langle n_i \rangle$. Most results were obtained
with a truncation number $m=200$. For several cases we examined also
$m=300$ and $m=400$ with almost indistinguishable results.

In principle, we are interested in the effect of oxygen depletion
which implies that CuO chains are cut at oxygen sites. These
depleted oxygens could be modeled by just imposing open boundary
conditions on chains with odd number of sites with the two end sites
corresponding to Cu-ions. However, it is technically convenient to
use chains with even number of sites which implies that one end
is occupied by an oxygen ion and the other end site by a copper
ion. The open end with an O ion would correspond to a nonmagnetic
impurity, for example a Zn ion replacing a Cu ion.\cite{nonmagnetic}
If one assumes that the Friedel oscillations from both ends 
do not have nonlinear superposition, i.e. both oscillation are
independent of each other, then the use of even number of sites
allows us the simultaneous study of both types of impurities in
a single chain.

Since, as said in the Introduction, Friedel oscillations appear
due to oxygen depleted sites acting as impurity centers in an 
otherwise metallic chain, then they will follow a power law behavior
predicted by Luttinger liquid theory,\cite{schulz}
\beq
n(r)-n_0 \sim
a_1\frac{\cos(2k_Fr +\phi_1)}{r^{(1+K_\rho)/2}}
+a_2\frac{\cos(4k_Fr +\phi_2)}{r^{2 K_\rho}},
\label{powerlaw}
\eeq
with $r$ the distance from the impurity. This expression can
be derived from the asymptotic form of the charge correlations 
and $K_\rho$ is the interaction dependent correlation 
exponent.\cite{fabrizio} Numerical studies have verified that FO
follow this form for the one-band Hubbard and $t$-$J$ 
models on chains and ladders\cite{white_friedel} and for the Kondo
lattice model on chains.\cite{kondofriedel} In the following, we
will fit our data using Eq. (\ref{powerlaw}). The complete fitting
procedure is detailed in the Appendix. The key point of this 
procedure is that we {\em simultaneously} fit the FO
starting from the left chain end (Cu site) and the one which starts
from the right chain end (O site).

\subsection{Quarter filling}

\begin{figure}
\begin{center}
\setlength{\unitlength}{1cm}
\includegraphics[width=6cm,angle=-90]{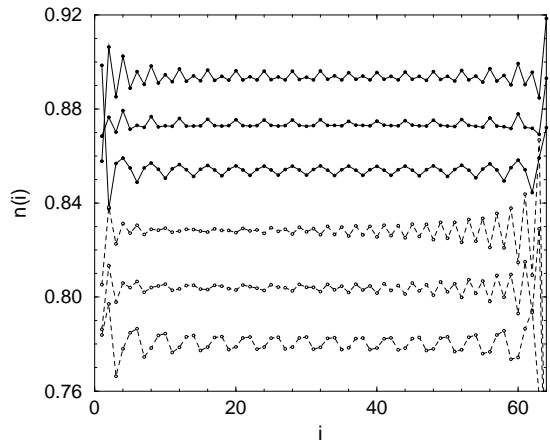}
\caption{Density at site $i$ for $L=64$, $n_h=0.5$, $U_d=8$, 
$U_p=2$, $U_{pd}=0$ and $\Delta=2$, 4 and 5 from bottom
to top. O sites: open symbols, Cu sites: filled symbols.
The curves have been arbitrarily shifted for the sake of clarity.
}
\label{osc05}
\end{center}
\end{figure}

We start by examining the electron density $n(i)$ as a function of
the position obtained by DMRG for $L=64$, $N_h=96$, which
corresponds to quarter filling, $n_h=0.5$,
as a function of $U_p$ and $\Delta$. At this filling,
the density modulation has period of four unit cell spacings for
the $2k_F$ component. This period can be clearly seen in the
Friedel oscillations shown in Fig. \ref{osc05}, ($U_p=2$) in both
Cu and O ions, at $\Delta=2$. These density oscillations extend
appreciably over the whole chain. For $\Delta=5$,
the density oscillations on both type of ions present a period
of two which corresponds to a $4k_F$ wave. Also shown are the
oscillations for $\Delta=4$, close to the crossover between
both regimes.

\begin{figure}
\begin{center}
\setlength{\unitlength}{1cm}
\includegraphics[width=6cm,angle=-90]{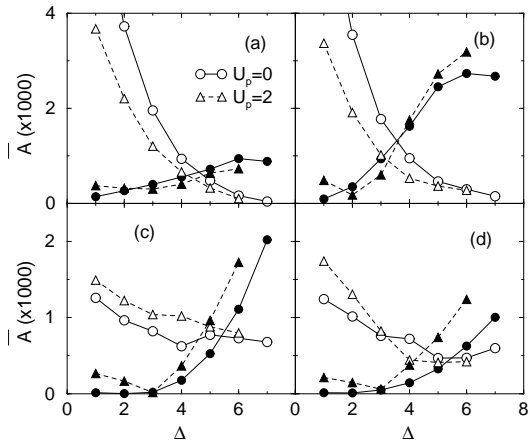}
\caption{$\Delta$ dependence of the   mean  amplitude of $2k_F$ 
(open symbols) and $4k_F$ (filled symbols) components for $U_d=8$,
$L=64$ and various values of $U_{p}$ at $n_h=0.5$.
(a) O density oscillations, O-impurity, 
(b) O density oscillations, Cu-impurity, (c) Cu density oscillations,
O-impurity, (d) Cu density oscillations, Cu-impurity, }
\label{ampn05}
\end{center}
\end{figure}

By applying the fitting procedure previously discussed, we are
able to obtain the   mean  amplitudes for the $2k_F$ and 
$4k_F$ components of the Friedel oscillations on Cu or on O ions
starting at site 1 (occupied by a Cu ion), corresponding to an
impurity on an O ion, and the one starting at site 64 (occupied by an
O ion), corresponding to an impurity on a Cu ion. These mean
amplitudes are defined by Eqs. (\ref{amp_L}) and (\ref{amp_R}).
The inclusion of the density
oscillation close to the ends introduces somewhat large 
numerical errors so we discarded the first three points near
the ends. The results are depicted in Fig.~\ref{ampn05}. The
error bars are about the size of the symbols used. By
discarding different number of
initial sites of course the amplitudes vary. However, their
crossing point does not change appreciably. Various features
can be rapidly appreciated: a larger amplitude
on O sites than on Cu sites for the dominant component; the
suppression of the $2k_F$ amplitude with increasing $U_p$
for O-site oscillations; larger (smaller) O (Cu) amplitudes for
Cu-impurity than for O-impurity.

\begin{figure}
\begin{center}
\setlength{\unitlength}{1cm}
\includegraphics[width=6cm,angle=-90]{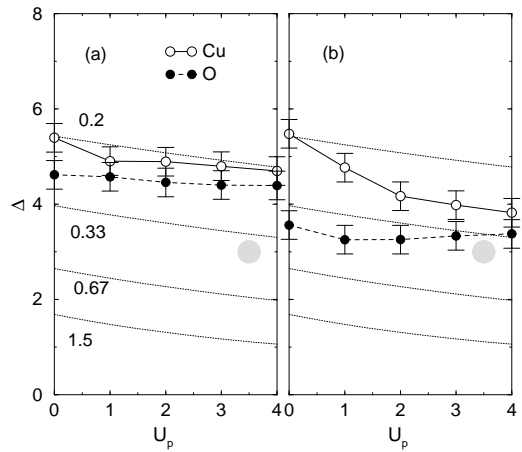}
\caption{Phase boundary between the $2k_F$ and $4k_F$ Friedel
oscillations at quarter filling for $U_d=8$, $L=64$, (a)
O-impurity; (b) Cu impurity. Open circles: boundary of 
Cu oscillations, filled circles: boundary for O oscillations. 
Dotted lines are lines of constant $J/t_{eff}$ of the effective 
one band $t$-$J$ model (see text). The gray circles correspond
approximately to the physical region of CuO chains.}
\label{phdiag_quar}
\end{center}
\end{figure}

The crossing of these   mean  amplitudes as $\Delta$ is increased
for a given value of $U_p$ is adopted as the 
crossing of the $2k_F$-dominated regime to the $4k_F$-dominated
regime. Other criteria could be eventually adopted but we found that
this criterion faithfully reproduces the change of oscillation
period which is the quantity measured in STM experiments.
The results can be summarized in the ``phase" diagram shown in
Fig.~\ref{phdiag_quar}. The region of $\Delta$ below the 
crossover corresponds to the $2k_F$-dominated regime while the
region of $\Delta$ above it corresponds to the $4k_F$-dominated
regime. The first remarkable feature is that the crossover in
the O-ions oscillations takes place at lower values of $\Delta$
than the crossover in the Cu-ions oscillations.

To help the interpretation of these results, we have overimposed
on this figure lines of constant $J/t_{eff}$ of an effective
one-band $t$-$J$ model with,\cite{brenig}
\beqa
J = \frac{4 t_{pd}^4}{\Delta^2} \left( \frac{1}{U_d} +
\frac{2}{2 \Delta + U_p} \right)
\eeqa
and
\beqa
t_{eff} \sim \frac{t_{pd}^2}{\Delta}
\eeqa
We have fixed the proportionality constant in such a way that the
physical point for the CuO chains corresponds to 
$J/t_{eff}=0.35-0.4$, taking into account some dispersion on the
values reported for the Coulomb interactions which are
approximately, $U_d\sim 8$ (the value adopted in the present study),
$U_p\sim 3-4$, $\Delta\sim 3$ (in units of $t_{pd}$). At a fixed
hole density, each line of constant $J/t_{eff}$ (indicated in
Fig.~\ref{phdiag_quar}) corresponds in the effective 1D $t$-$J$
model to a line of constant $K_{\rho}$, which
have been previously estimated.\cite{ogata} Then, it can be
appreciated in Fig.~\ref{phdiag_quar} that the $2k_{F}$/$4k_{F}$ 
boundaries are roughly parallel to these lines with the largest 
deviation corresponding to the O oscillations, Cu impurity. 
In addition, as it can be seen in 
Fig.~\ref{phdiag_quar}(a), for the physical region corresponding to
the CuO chains (indicated with a gray circle), the
FO generated by oxygen depletion have $2k_{F}$ modulation in
agreement to experimental results near quarter filling.\cite{maki}
In the case of Cu impurity (Fig.~\ref{phdiag_quar}(b)),
although the $2k_{F}$/$4k_{F}$ boundary is shifted to lower values
of $\Delta$, the FO are predicted to have also a $2k_{F}$ character
in the physical region.

\subsection{$n_h=0.875$ filling}

While in the undoped compound, YBa$_2$Cu$_3$O$_7$, the chain filling
is $n_h=0.5$, upon oxygen depletion it increases toward
half-filling, which is reached at $\delta \approx 0.36$
(Ref.~\onlinecite{cava}).
Hence, in this subsection we consider the case of $n_h=0.875$.
Although increased O depletion would produce on average shorter 
CuO segments, in order to reduce the number of varying parameters
and to facilitate the comparison, we study the same chain
length as in the previous subsection, i.e. $L=64$. The
filling $n_h=0.875$ then corresponds to $N_h=72$ holes.
In this case, $2k_{F}=7\pi/8$ and $4k_{F}=\pi/4$.

\begin{figure}
\begin{center}
\setlength{\unitlength}{1cm}
\includegraphics[width=6cm,angle=-90]{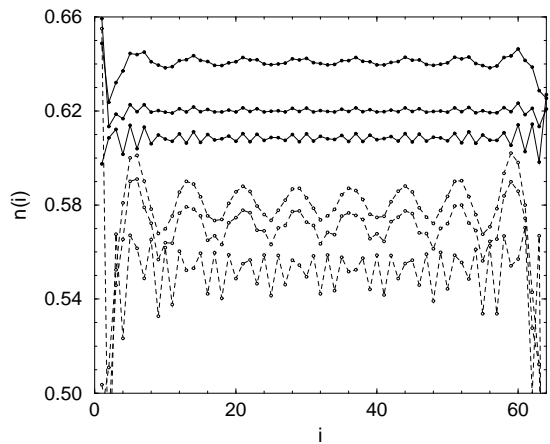}
\caption{Density at site $i$ for $L=64$, $n_h=0.875$, $U_d=8$, 
$U_p=0$, $U_{pd}=0$ and $\Delta=1$, 3 and 5 from bottom
to top. O sites: open symbols, Cu sites: filled symbols.
The curves have been arbitrarily shifted for the sake of clarity.
}
\label{osc0875}
\end{center}
\end{figure}

Let us start by examining in real space density oscillations
corresponding to this filling. They are illustrated in 
Fig.~\ref{osc0875}. The oscillations for small $\Delta$,
corresponding to $2k_{F}$ can be regarded as period 2
oscillations with kinks, while the oscillations at large
$\Delta$, corresponding to $4k_{F}$ have a well defined 
period 8 unit cell spacings.

\begin{figure}
\begin{center}
\setlength{\unitlength}{1cm}
\includegraphics[width=6cm,angle=-90]{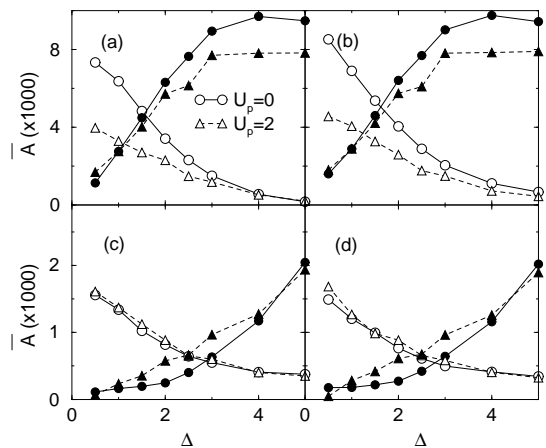}
\caption{Same as Fig.~\ref{ampn05} but for 
$n_h=0.875$.}
\label{ampn0875}
\end{center}
\end{figure}

As we did in the previous subsection, we have fitted the density
oscillations for both Cu and O ions and the results for the 
  mean  amplitudes of the $2k_{F}$ and $4k_{F}$ components are 
shown in Fig.~\ref{ampn0875}.  By comparing with the previous
results at quarter filling, it can be concluded that the amplitudes
of the dominant component are in general much larger in this case of
$n_h=0.875$. Again O oscillations have in general larger amplitudes
than the Cu ones. It can be also observed that the effect of $U_p$
is relatively small in the present case. Also there are
virtually no differences between the cases of O-impurity and
Cu-impurity.

\begin{figure}
\begin{center}
\setlength{\unitlength}{1cm}
\includegraphics[width=6cm,angle=-90]{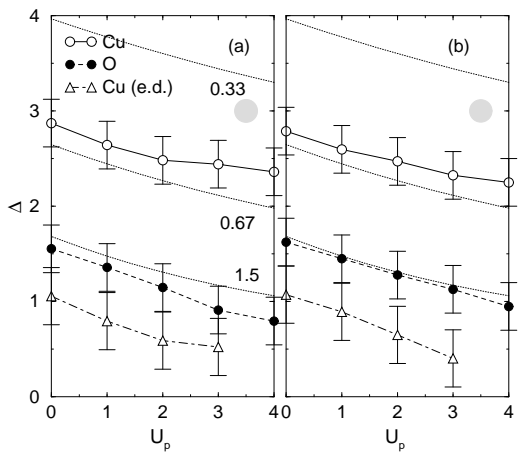}
\caption{Phase boundary between the $2k_F$ and $4k_F$ Friedel
oscillations for $U_d=8$, $L=64$, $n_h=0.875$. (a)
O-impurity; (b) Cu impurity. Open circles: boundary of 
Cu oscillations, filled circles: boundary for O oscillations. 
Open triangles correspond to Cu oscillations in electron doped,
$n_e=0.875$, chains.
The gray circles correspond
approximately to the physical region of CuO chains.}
\label{phdiag_half}
\end{center}
\end{figure}

Again, the crossing points between  $2k_F$ and $4k_F$ components
of the   mean  amplitudes 
lead to the ``phase" diagram shown in Fig.~\ref{phdiag_half}.
As anticipated by the previous figure, the boundary between the
$2k_F$ (low $\Delta$) and the $4k_F$ (high $\Delta$) regions
has been considerably shifted to lower values of $\Delta$ 
with respect to the quarter filled case. As for $n_h=0.5$, these
boundaries also follow approximately the lines of constant 
$K_{\rho}$ of the effective 1D $t$-$J$ model. However, in contrast
with that case, there is much less difference between the boundaries
for O depletion-induced FO (Fig.~\ref{phdiag_half}(a))
and the boundaries for the FO induced by nonmagnetic substitution of
Cu ions (Fig.~\ref{phdiag_half}(b)). In any case, as in the quarter
filled case, the most remarkable feature is again that the 
$2k_F$ to $4k_F$ crossover for oxygen FO takes place at smaller
values of $\Delta$ than the one for Cu FO.
Notice that for this filling the physical region of parameters
(gray circles) falls now on the $4k_F$ side of the diagram.

Finally, by further oxygen depletion, $\delta > 0.36$,
the chain would formally go into the electron doped region.
The $2k_F$/$4k_F$ boundary between Friedel oscillations
on Cu ions in the electron doped, $n_e=0.875$, system is also
included in Fig.~\ref{phdiag_half}. This result confirms the
trend suggested by the cases $n_h=0.5$ and $n_h=0.875$, that
by reducing the number of holes this boundary shifts to
lower values of $\Delta$ for a fixed $U_p$. On the other hand,
the occupancy on O sites is very low and we observed $2k_F$
oscillations for all the values of parameters examined. It is
reasonable to speculate that by further reducing the number of
holes the presence of $4k_F$ oscillations on Cu ions will
eventually disappear as well, i.e. the effects of strong
electron correlations are expected to be more important close to
half-filling.

\subsection{Global properties}

\begin{figure}
\begin{center}
\setlength{\unitlength}{1cm}
\includegraphics[width=6cm,angle=-90]{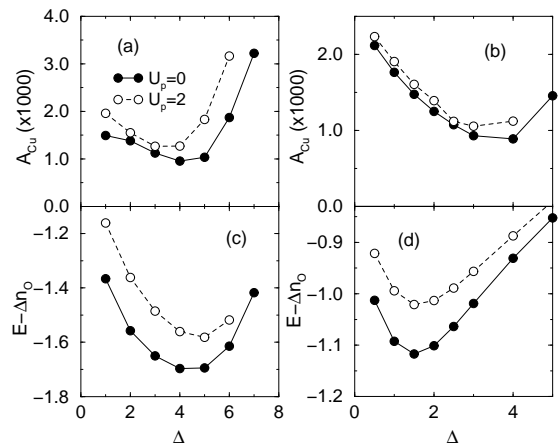}
\caption{(a) Average amplitude of Friedel oscillations on Cu sites 
for $n_h=0.5$. (b) Same for $n_h=0.875$. (c) Corrected ground state
energy for $n_h=0.5$. (d) Same for $n_h=0.875$.}
\label{global}
\end{center}
\end{figure}

Let us examine some global properties like the total ground state 
energy $E_0$ and the average amplitude of FO.
Since this energy is monotonically increasing with $\Delta$, for all
the other parameters fixed, it is more meaningful to analyze the
quantity obtained by subtracting from $E_0$ the quantity
$\Delta n_O$ where $n_O$ is the average filling on O sites.
The average of the amplitude of the FO on Cu
sites is defined as,
\beqa
A_{Cu}=\frac{1}{L} \sum_{i} |n(i)-n_{Cu}|
\label{avampCu}
\eeqa
where the sum runs over Cu sites and $n_{Cu}$ is the average filling
on Cu sites. To be consistent with the fitting procedure and hence
to allow comparison with Figs.~\ref{ampn05} and \ref{ampn0875} we
discarded the first three sites close to both chain ends. The
resulting quantity is shown in Fig.~\ref{global}(a),(b)
for $n_h=0.5$ and $n_h=0.875$ respectively. It can be noticed that it
presents a minimum at a value of $\Delta$ close to the 
$2k_{F}$/$4k_{F}$ boundary of Cu FO at a given value of $U_p$. As it
can be seen in Fig.~\ref{global}(c),(d), the corrected energy, 
$E_0-\Delta n_O$ also presents a minimum at approximately the same
values of $\Delta$ for $n_h=0.5$. At $n_h=0.875$, the minimum of
this quantity is located at a lower value, closer to the
$2k_{F}$/$4k_{F}$ boundary of oxygen FO.

\section{Discussions and conclusions}
\label{discussion}

The two most important results of the present study, which can be
deduced from Figs. ~\ref{phdiag_quar} and ~\ref{phdiag_half} are:
(i) the $2k_{F}$
to $4k_{F}$ crossover takes place at smaller values of
$\Delta$ for given $U_p$ as doping is reduced from quarter 
filling to half filling, and (ii) for hole doped chains this
crossover takes place at smaller $\Delta$ for the density
oscillations on O ions than for the ones on Cu ions. Let us
discuss these two features in detail.

In the first place, the fact that the $2k_{F}$/$4k_{F}$ boundaries
quite likely correspond to lines of constant $K_{\rho}$ of an
effective 1D $t$-$J$ model suggests
that both the approximate mapping to this model and the
whole fitting procedure are physically correct. One should notice 
that this effective one band model is obtained essentially by
projecting out the oxygen sites, i.e. is a model valid mostly
for Cu ions. Besides, it becomes less valid as one moves away from
half filling. It is not 
obvious then that also the boundaries for oxygen oscillations
follow approximately the lines of constant $J/t_{eff}$
although with some important deviations, specially at quarter
filling. The dependence of $K_{\rho}$ with $J/t$ and hole density
for the 1D $t$-$J$ model\cite{ogata}, is consistent with the
shifting of the Cu $2k_{F}$/$4k_{F}$ boundary to lower values of 
$\Delta$ as hole density $n_h \rightarrow 1$, and in turn this
is consistent with the intuitive notion that strongly correlated
electrons regimes become more important close to half-filling.

However, this mapping does not explain by itself why in the case
of hole doping, the $2k_{F}$/$4k_{F}$ boundaries for O oscillation
occur at smaller values of $\Delta$ for any given $U_p$, i.e.
why the O-subsystem enters in the strong correlation regime 
at smaller interactions than the Cu-subsystem. To understand
this feature, it would be necessary to analyze effective models
obtained by projecting out Cu sites and retaining O sites.
These effective models, the so-called ``spin-fermion" or
Kondo-Heisenberg models,\cite{brenig} are considerably more
complicated and related available results are
limited.\cite{sikkema}
For hole doped cuprates, i.e. with holes in excess of the number
of Cu ions,
it can be assumed that Cu ions are mostly single occupied, and in
this case our results suggest that the $2k_{F}$ to $4k_{F}$
crossover is ``driven" by the oxygen oscillations. That is, the
O subsystem is more susceptible to strong correlations than the
Cu subsystem.
On the other hand, for the case of electron doped chains, i.e.
with fewer holes than Cu ions, the $2k_{F}$ to $4k_{F}$ crossover
takes place at small $\Delta$ for Cu density oscillations while
for O FO have $2k_{F}$ in the range of parameters studied, and
in this sense, the crossover seems ``driven"
by the Cu ions. Hence, the nearly empty O band in electron doped
chains is symmetric to the half-filled Cu band in hole doped
cuprates. 

There are two factors affecting the magnitude of the mean
amplitudes of the $2k_{F}$ and $4k_{F}$ components: their 
amplitude at the origin, and their power law exponents. The
amplitudes at the origin ($A_2$, $A_3$, $B_2$, $B_3$, in Eqs.
(\ref{left}) and (\ref{right})) may depend on the formation
of bound states between holes and impurities. With respect to the
power law exponents, one could assume that the FO on Cu and
on O sites are determined by two different $K_{\rho}$ since they
obey different effective models in addition of having different
hole densities.
Unfortunately, although the fitting to the oscillations using
Eqs.~(\ref{left}), (\ref{right}), (\ref{sum}) are very good, the
determination of $K_{\rho}$ from the fitted values is very noisy to
allow reliable conclusions. In this sense, an independent study of
{\em periodic} chains, with a careful study of finite size effects,
would be necessary to confirm this possibility.

{\em Order of the crossover}. It seems to be a first order in the
sense that there are no other oscillation wavenumber between
$2k_{F}$ to $4k_{F}$. The evidence for this is rather indirect since
our method of fitting does not make room for variable wavenumber,
and a Fourier (or windowed Fourier) transform would mix the
Friedel oscillations starting form the left and right ends of the
open chain. The indication of a first order type of crossover
comes from the fact that the overall average amplitude of the
density oscillations are minimum at the crossing point 
(Figs.~\ref{osc05}, ~\ref{osc0875} and ~\ref{global}(a),(b)). This
suggests that the system becomes ``frustrated" at the crossing point
due to the competition of $2k_{F}$ and $4k_{F}$ modulations being
unable to develop a modulation at another wavenumber. The behavior
of the corrected ground state energy, $E_0 -\Delta n_O$,
(Fig.~\ref{global}(c),(d)) is 
consistent with this interpretation: its minimum is located
close to the point where $A_{Cu}$ is also minimum presumably because
it gains energy due to delocalization. Hence, this quantity also
suggests that there are no FO with a modulation intermediate between
$2k_{F}$ and $4k_{F}$.

{\em Predictions}. The main prediction of the present study is that 
CuO chains in YBa$_2$Cu$_3$O$_{7-\delta}$ would undergo a 
$2k_{F}$ to $4k_{F}$ crossover in the Friedel oscillations
induced by oxygen depletion or by nonmagnetic substitution of
Cu ions as doping moves from quarter to half filling and 
eventually into the electron doped region. Additionally we
predict that experiments like STM, probing O sites, would 
detect a $4k_{F}$ modulation while experiments such as neutron
scattering would still see a $2k_{F}$ modulation on Cu sites.
More generally, the modulation of FO can be
considered as a sensible tool to detect the sometimes subtle
presence of strong electron correlations in quasi-1D systems.
Finally, if the observed modulations in CuO chains are 
essentially Friedel oscillations and the presence of 
superconductivity on the planes plays a minor role, then similar
modulations should be observed in the non-superconducting
compound PrBa$_2$Cu$_3$O$_{7}$ (Ref.~\onlinecite{mizokawa}).

\begin{acknowledgments}
This work was supported by a Grand-in-Aid for Scientific Research on
Priority Areas  and the NAREGI Nanoscience Project from MEXT and
CREST. One of authors (M. M.) acknowledges support of 21st. Century
COE program.
The use of supercomputers and friendly technical assistance at the 
Center for Computational Materials Science, IMR, Tohoku University,
is also gratefully acknowledged.
\end{acknowledgments}

\appendix
\section{Fitting procedure}
\label{fit_proc}

In general, the oscillation starting from the Cu-edge is different to
that from the O-edge. The total oscillation, as obtained numerically,
is the superposition of both oscillations. Assuming these oscillation
follow the Luttinger power law expressions,
\beq
L(r) \equiv
A_0+ A_1\frac{1}{r}
+A_2\frac{\cos(2k_F r +\phi_1)}{r^{\gamma_1}}
+A_3\frac{\cos(4k_F r +\phi_2)}{r^{\gamma_2}},
\label{left}
\eeq
and
\beqa
R(r) &\equiv&
B_0+ B_1\frac{1}{L-r}
+B_2\frac{\cos(2k_F(L-r) +\varphi_1)}{(L-r)^{\eta_1}}
\nonumber\\
&+&B_3\frac{\cos(4k_F(L-r) +\varphi_2)}{(L-r)^{\eta_2}},
\label{right}
\eeqa
where $L$ is the system size, then our fitting function will be,
\beqa
S(r) \equiv L(r) + R(r)
\label{sum}
\eeqa
with $n_0<r<L-n_0$. $n_0$ is the number of sites from each edge
that are cut out to avoid wild oscillations.
Below, the constant terms, $A_0$ and $B_0$,  are neglected, since a
uniform component is subtracted before the fitting process. 

The mean amplitudes of the $2k_F$ and $4k_F$ components of the
Friedel oscillations starting from the
left and from the right edges of the chain are defined as,
\beqa
\bar{A}_{2k_F,L}=\frac{A_2}{L-2 n_0}\int_{n_0}^{L-n_0}\rmd r 
\frac{1}{r^{\gamma_1}} \nonumber  \\
\bar{A}_{4k_F,L}=\frac{A_3}{L-2 n_0}\int_{n_0}^{L-n_0}\rmd r 
\frac{1}{r^{\gamma_2}}
\label{amp_L}
\eeqa
and
\beqa
\bar{A}_{2k_F,R}=\frac{B_2}{L-2 n_0}\int_{n_0}^{L-n_0}\rmd r 
\frac{1}{r^{\eta_1}} \nonumber  \\
\bar{A}_{4k_F,R}=\frac{B_3}{L-2 n_0}\int_{n_0}^{L-n_0}\rmd r 
\frac{1}{r^{\eta_2}}
\label{amp_R}
\eeqa
respectively.

In the fitting process, Eq.~(\ref{sum}) is rewritten as,
\beqa
S(r_i) &\equiv&
A_1\frac{1}{r_i}
+A'_2\frac{\cos(2k_Fr_i)}{r_i^{\gamma_1}}
+A''_2\frac{\sin(2k_Fr_i)}{r_i^{\gamma_1}}\nonumber\\
&+&A'_3\frac{\cos(4k_Fr_i)}{r_i^{\gamma_2}}
+A''_3\frac{\sin(4k_Fr_i)}{r_i^{\gamma_2}}
+B_1\frac{1}{L-r_i}\nonumber\\
&+& 
B'_2\frac{\cos(2k_F(L-r_i))}{(L-r_i)^{\eta_1}}
+B''_2\frac{\sin(2k_F(L-r_i))}{(L-r_i)^{\eta_1}}\nonumber\\
&+&B'_3\frac{\cos(4k_F(L-r_i))}{(L-r_i)^{\eta_2}}\nonumber\\
&+&B''_3\frac{\sin(4k_F(L-r_i))}{(L-r_i)^{\eta_2}}.
\label{sum2}
\eeqa

\end{document}